\documentclass[10pt,twocolumn,superscriptaddress,aps,showpacs,preprintnumbers,amsmath,prb]{revtex4-1}
\usepackage{graphicx}
\usepackage{dcolumn}
\usepackage{bm}
\usepackage{color}
\usepackage{hyperref}
\usepackage{float}
\usepackage[normalem]{ulem}

\newcommand{\mev}{\text{meV}}

\usepackage{graphicx}
\usepackage[normalem]{ulem}
\usepackage{color}
\usepackage{amsmath}
\usepackage{enumitem}

\newcommand{\pd}[2]{\frac{\partial #1}{\partial #2}}
\renewcommand{\S}{\bm{S}}

\begin{document}
\author{David Sanz Ruiz}
\affiliation{Departamento de Física Aplicada, Universidad de Alicante, 03690 Alicante, Spain}
\author{David Soriano}
\email[Email:]{david.soriano@ua.es}
\affiliation{Departamento de Física Aplicada, Universidad de Alicante, 03690 Alicante, Spain}
\title{Enhancement of topological magnon-driven spin currents through local edge strain in CrI$_3$ nanoribbons}

\begin{abstract}
This work describes topological magnon transport in zigzag CrI$_3$ nanoribbons (ZNR) in presence of edge strain. Exchange coupling terms under strain are obtained from first-principles calculations, and the topological properties are introduced \emph{via} second-neighbor Dzyaloshinskii-Moriya interactions. The magnon Hamiltonian is calculated using linear spin-wave theory and the Holstein-Primakoff transformation. Then, we use non-equilibrium Green's function method to calculate the spin-wave-generated currents in ribbons with different edge strain. Our calculations show the formation of strongly localized edge topological magnons within the gap for DMI values slightly higher than the ones reported experimentally and in the presence of a tensile edge strain of the order of 3\%. The magnon-mediated topological spin transport calculations shows an increase of the spin current and characteristic decay length in tensile-strained CrI$_3$ nanoribbons compared with unstrained ones. Our findings demonstrate that straintronics provides a powerful route to harness and control topological magnons in two-dimensional magnetic materials.          
\end{abstract}

\maketitle
\section{Introduction}
Magnetic two-dimensional (2D) materials have strongly revolutionized the fields of spintronics, allowing us to explore vibrant spin phenomena in few-nanometer-thick van der Waals heterostructures, ranging from ultrathin van der Waals magnetic tunnel junctions\cite{Klein2018, Wang2018} to topological superconductivity\cite{Kezilebieke2020}. A very appealing application of these materials resides in the transmission of information {\it via} electron spins by spin-waves rather than electric currents, also known as insulator spintronics or low-power spintronics\cite{Chumak2015,Samuel2025}. However, despite their considerable technological potential, ferromagnetic two-dimensional insulators have seen limited progress toward practical implementation in magnon-assisted spin transport. One of the primary limitations arises from the difficulty of detecting spin waves in ultrathin magnetic layers, where the reduced magnetic volume leads to significantly weaker signals compared to bulk materials such as yttrium iron garnet(YIG).\cite{Wei2022}.

Another notable property of some two-dimensional magnetic materials, including CrI$_3$\cite{huang2017layerCrI3}, is the appearance of topological magnon bands generated by spin–orbit–coupling–induced Kitaev\cite{Aguilera2020} or second-neighbor Dzyaloshinskii–Moriya\cite{Costa2020} interactions in hexagonal lattices. Neutron scattering experiments on bulk CrI$_3$ show a gapped spin-wave dispersion at K points\cite{Chen2018}, very similar to its fermionic homologous in hexagonal lattices\cite{Kane2005}. Very important questions in this regard are how topological magnons can be detected in 2D ferromagnets\cite{zhang2024} and which mechanisms can facilitate their detection.

Motivated by these questions and previous works on magnon straintronics\cite{Esteras2022}, we use local edge strain to engineer the magnon spectrum of zigzag CrI$_3$ nanoribbons so that the topological magnon bands lie closer to thermally accessible energies near the magnon gap, thereby increasing their potential impact on spin transport and their experimental detectability. The paper is organized as follows: in section \ref{sec2} , we describe the methodology used throughout the manuscript (spin Hamiltonian, Holstein-Primakoff, non-equilibrium magnon transport, etc.). In section \ref{sec3}, we show the spin-wave dispersion of CrI$_3$ nanoribbons for different widths and in presence and absence of second-neighbor Dzyaloshinskii-Moriya interactions (DMI). We demonstrate how traversal strain can tune the frequency of topological edge magnons, bringing them into the gap. In section \ref{sec4}, we study the spin transport in strained CrI$_3$ nanoribbons mediated by spin-waves using the non-equilibrium Green's function formalism as previously reported by R{\"u}ckriegel \emph{et al.}\cite{RuckriegelDuine2018-cx}. Finally, in section \ref{sec5} we discus the results and report the main conclusions.

\section{Methodology}
\label{sec2}

\subsection{Strain-dependent exchange interactions from DFT}

Density Functional Theory (DFT) calculations were performed using the Quantum Espresso \emph{ab initio} package \cite{Giannozzi2009, Giannozzi2017} to investigate exchange interactions in monolayer CrI$_3$ under various homogeneous strain conditions. The primary objective was to determine the exchange parameters as a function of strain, allowing for the generalization of this relationship to the non-homogeneous strain profiles proposed for engineering magnon states.

We employed self-consistent spin-polarized DFT calculations within the Generalized Gradient Approximation (GGA), using the Perdew-Burke-Ernzerhof (PBE) exchange-correlation functional \cite{DalCorso2014} and the Projector Augmented Wave (PAW) method \cite{Blochl1994}. Since we are interested in Heisenberg exchange terms, spin-orbit coupling (SOC) was excluded from these specific calculations. To account for electron-electron correlations and stabilize the ferromagnetic ground state of the CrI$_3$ monolayer, we applied the DFT+$U$+$J$ scheme, incorporating both on-site Hubbard ($U$) and Hund's exchange ($J$) mean-field interactions. Specifically, we adopted values of $U=2.7$ eV and $J=0.7$ eV from ref.\cite{Lado2017-jg}. For all the calculations, we used an $8\times8\times1$ $k$-point mesh, an energy convergence threshold of $10^{-9}$ Ry, and a lattice parameter of $a = 7.01$ \AA. 

We investigated the effect of in-plane biaxial strain in monolayer CrI$_3$ by distorting the lattice vectors within a range of $\pm3\%$. For each strain configuration we calculated total energies of ferromagnetic and antiferromagnetic orders to establish a functional relationship between the exchange interactions and lattice distortions. The exchange interaction parameters were extracted by mapping the total energy differences between the ferromagnetic (FM) and antiferromagnetic (AFM) ground states to a nearest-neighbor spin Heisenberg model of the form:
\begin{equation}
    H = -\frac{1}{2}\sum_{ij} J_{ij} \mathbf{S}_i \cdot \mathbf{S}_j,
\end{equation}
where $J_{ij}$ represents the exchange coupling between nearest neighbors and $S_i,S_j$ are spin vectors at $i$ and $j$ lattice sites. The general method to obtain exchange parameters directly from the total energies of \textit{ab initio} calculations is to compare the FM and AFM energies (specifically a N\'eel AFM state), using the relation:
\begin{equation}
    J = \frac{E_{\text{AFM}} - E_{\text{FM}}}{2zS^2},
\end{equation}
where $z$ is the neighbor coordination number. For the monolayer calculations used to extract the bulk parameter, $z=3$.

\begin{figure}[!t]
    \centering
    \includegraphics[width=1.0\linewidth]{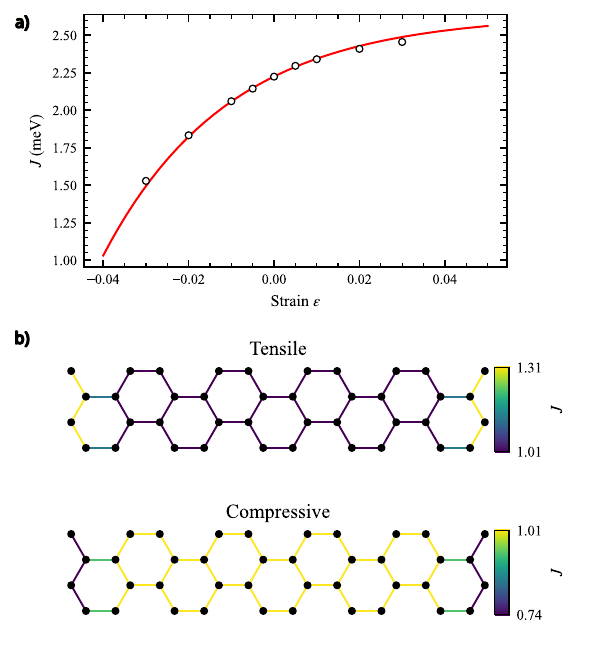}
    \caption{(a) Nearest-neighbor exchange interactions obtained from \textit{ab initio} DFT calculations (black dots) for a monolayer under homogeneous biaxial strain. The strain-dependent exchange coupling is interpolated using the phenomenological model $J(\varepsilon) = J_0(a - be^{c\varepsilon})$ (red curve), with the unstrained exchange parameter $J_0 = 2.2$ meV and  fitted parameters $a = 1.1987$, $b = 0.1874$, and $c = 33.9837$. The strain parameter is defined as $\varepsilon = a'/a - 1 = a'_{0}/a_{0} - 1$, where $a'$ ($a'_{\text{0}}$) and $a$ ($a_{\text{0}}$) denote the strained and unstrained lattice parameters (nearest-neighbor distances), respectively. (b) Spatial distribution of nearest-neighbor exchange interactions obtained by evaluating the fitted model locally in the presence of non-uniform strain field in units of $J_0$. The system is parametrized by a width $W = 10$, a smoothing length $\tau = a_{\text{0}}/4$ and strain amplitudes $\alpha_{str} = -0.3$, $0.6$ \AA. } 
    \label{fig:xc_strain_dft}
\end{figure}

The results of these calculations are presented in Fig.~\ref{fig:xc_strain_dft}. For the unstrained system we find $J=2.2\,\mev$, which is in good agreement with earlier reported results\cite{Lado2017-jg}. We observe an asymmetric response to the applied deformations: compressive strain has a stronger impact on the exchange interactions, decreasing the coupling, whereas tensile strain enhances the exchange interactions following a more moderate profile. These strain-dependent tendencies are consistent with recent first-principles studies \cite{2022Wei, 2023SoenenMaartenMilo}. 

We modeled the strain dependence of the magnetic exchanges between neighboring sites, $J(\varepsilon)$, using the exponential function:
\begin{equation} \label{eq:interpolated-model}
    J(\varepsilon) = J_0 \left( a - b e^{c\varepsilon} \right),
\end{equation}
where the parameters $a$, $b$, and $c$ are fitted to the DFT data as shown in Fig.~\ref{fig:xc_strain_dft}(a). The strain parameter is defined by the local relative change in nearest-neighbor distances, $\varepsilon = \frac{a'_0}{a_0}- 1$.

This functional relationship allows us to introduce a non-uniform strain profile characterized by distortions localized at the edges of the ribbon. This approach exploits the imbalance in the coordination number between edges and bulk of nanoribbons, which enables the tuning of the magnons of the system. Specifically, we propose the following displacement field:
\begin{equation}
    \Delta u_x = \alpha_{\text{str}} \left( -e^{-(x-x_L)/\tau_{\text{str}}} + e^{(x-x_R)/\tau_{\text{str}}} \right),
\end{equation}
where we introduce the parameter $\alpha_{\text{str}}$ that controls the amplitude of the lattice deformations. and is given in units of \AA. The maximum strain value that will be applied for a fixed $\tau=a_0/4$ is given by $\sim\frac{\alpha_{\text{str}}}{2a_0}$. 
$x_L$ and $x_R$ denote the left and right edge positions of the ribbons, and $\tau$ (in units of $a_0$) modulates the spatial decay of the deformations.  

Figure \ref{fig:xc_strain_dft}(b) illustrates the resulting exchange interactions under the proposed strain profile. Due to the asymmetric dependence of $J$ with respect to $\varepsilon$ (see Eq.~\ref{eq:interpolated-model}), compressive strains exert a stronger influence on magnon dynamics than tensile strains. We restrict the applied deformation field to nearly the ranges explored in Fig.~\ref{fig:xc_strain_dft}(a).

\subsection{Spin-wave Hamiltonian of CrI$_3$ nanoribbon}

In this work, we focus on the isotropic exchange interactions and its dependence on lattice deformations, extracted from DFT calculations. For this aim, we propose a Heisenberg model with second-neighbor Dzyaloshinkii-Moriya Interactions (DMI) to describe the spin dynamics in CrI$_3$ nanoribbons. The system is modeled using the following spin Hamiltonian: 
\begin{equation}\label{eq:general_heis_ham11}
\begin{split}
    H = &-\sum_{\langle i j \rangle} J_{ij}(\varepsilon)\bm{S}_i\cdot \bm{S}_j
    - \sum_{\langle i j \rangle} \lambda S^z_iS^z_j \,, \\
    &+\sum_{\langle\langle i j \rangle\rangle}\bm{D}_{ij} \cdot (\bm{S}_i\times\bm{S}_j)
    - \sum_i \bm{h}_i\cdot\bm{S}_i\,.
\end{split}
\end{equation}
The first term is the nearest neighbor isotropic term with $\bm{S}_i$ being the spin operator associated with the $i$ site and $J(_{ij}\epsilon)$ the strain-dependent Heisenberg exchange coupling, where the strain value is computed as the local change in the nearest-neighbor bond distance between sites $i$ and $j$ as $\varepsilon = \tfrac{d_{ij}}{d_0}-1$, with $d_0$ the unstrained bond distance and $d_{ij}$ the strained one.  The second term contains the anisotropic exchange $\lambda = 0.1\,\text{meV}$ \cite{Lado2017-jg} which couples out-of-plane spins. The third term is the second-neighbor DMI which reflects the effect of the spin-orbit coupling (SOC) induced by the iodine ligands. $\bm{D}_{ij} = D \nu_{ij}$ is the antisymmetric DMI which captures clockwise hopping paths and encodes the system's topological characteristics\cite{Owerre2016-ho}. Last terms includes the effect of an external out-of-plane magnetic field. 

The low energy spin excitations associated with the spin Hamiltonian in Eq.\ref{eq:general_heis_ham11} are given in the form of quantized spin-waves, i.e., magnons. We use Linear Spin-Wave Theory (LSWT) and the Holstein-Primakoff (HP) transformations to obtain the spin-wave Hamiltonian \cite{PhysRev.58.1098}:
\begin{equation}
\begin{split}
    &S^+_i \approx \sqrt{2S}a_i \,,\\
    &S^-_i \approx \sqrt{2S}a^{\dagger}_i \,,\\
    &S^z_i = S - a^{\dagger}_ia_i \,.
\end{split}
\end{equation}
where $S^{\pm}_i = S^x_i\pm iS^y_i$ are the ladder spin operators at $i$ site
 and $a_i^{(\dagger)}$ are the annihilation (creation) HP magnon operators. We apply HP only up to quadratic order and neglect higher ones related with magnon-magnon scattering interactions. The quadratic form of the LSWT Hamiltonian in real space $H$  is expressed as,
\begin{equation}\label{eq:lswt-ham}
\begin{split}
    H &= \sum_{ij}
        \left\{
            \delta_{ij}\bigl[\sum_{n}S(J_{in}(\epsilon)+\lambda_{in})
            +h^{\text{ext}}\bigr] 
            + T_{ij}
        \right\} a^{\dagger}_ia_j\,,\\
    &T_{ij} = -S\left(J_{ij}(\epsilon)+ iD\nu_{ij}\right)\,.
\end{split}
\end{equation}

 The strain-dependent exchange interaction is given by $J_{ij}(\varepsilon)$ for nearest-neighbor sites, while $\nu_{ij}=\pm1$ determines the sign of the Dzyaloshinskii–Moriya interaction between next-nearest neighbors. The latter term is responsible for the emergence of nontrivial topological features in the magnon spectrum. The parameter $\lambda_{ij}$ denotes the anisotropic exchange interaction between nearest neighbors and $h^{\mathrm{ext}}$ is an external magnetic field applied to stabilize the magnetic order.

 By applying periodic boundary conditions and performing a Fourier transform, the Hamiltonian in reciprocal space becomes,
 \begin{equation}\label{eq:Hij_momentum}
    \begin{split}
    &H = \\
    &\quad \sum_{k}\sum_{ij}
    \bigl\{\delta_{ij}\bigl[\sum_{n} S(J_{in}(\epsilon)+\lambda_{in})+h^{\text{ext}}\bigr]
    +T_{ij}(k)\bigr\}a^{\dagger}_{k,i}a_{k,j}\,,\\
    &T_{ij}(k) = -S\left(J_{ij}(\epsilon)f_{ij}^{(1)}(k) + iD\nu_{ij}f_{ij}^{(2)}(k)\right)\,,
\end{split}
\end{equation}
where now $i$ and $j$ indices refer to sites within the unit cell, $f^{(1)}_{ij}(\mathbf{k}) = \sum_{\boldsymbol{\delta}_{ij}} 
e^{i\mathbf{k}\cdot \boldsymbol{\delta}_{ij}}$ and $f^{(2)}_{ij}(\mathbf{k}) = \sum_{\boldsymbol{\Delta}_{ij}} 
e^{i\mathbf{k}\cdot \boldsymbol{\Delta}_{ij}}$ denote the geometric structure factors resulting from the lattice vectors connecting nearest and next-nearest neighbors, respectively. It is worth noting that in Eq.\ref{eq:lswt-ham} and Eq. \ref{eq:Hij_momentum} the on-site terms depend only on the local coordination number. In particular, for a bulk site in the nanoribbon $\epsilon_\text{bulk} = 3JS$, whereas for an edge site $\epsilon_\text{edge} = 2JS$.

Although the magnon Hamiltonian is structurally very similar to the standard tight-binding model of graphene \cite{castro2009electronic}, the magnon on-site terms originate from the exchange interaction and therefore depend on the local coordination number of each site. This difference between edge and bulk on-site terms leads to the disappearance of the flat band typically expected for zigzag graphene nanoribbons, as previously reported \cite{aguilera2020topological}.

\subsection{Non-equilibrium Green's function method for spin-wave transport}

\begin{figure}[ht]
    \centering
    \includegraphics[width=\linewidth]{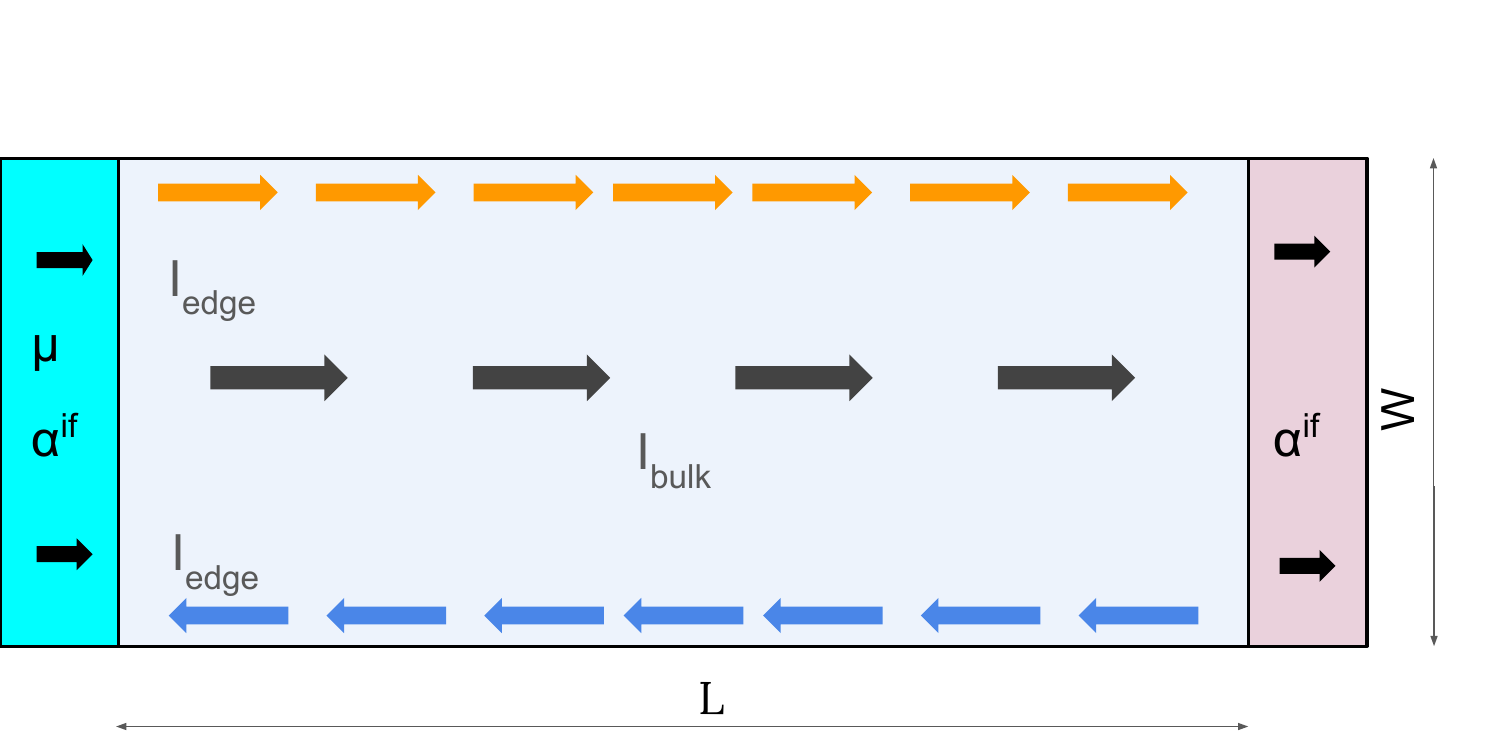}
    \caption{Scheme of the transport device used for the simulation of magnon spin transport in nanoribbons. $\bm{\mu}$ represents the spin accumulation on the injecting lead on the left that couples to the spins of the ZNR manifested as an effective magnetic field. $\alpha^{\text{if}}$ represents the interfacial Gilbert damping that appears due to the contacts. Magnon modes excites at this edge propagate through the ZNR to the right side where the current is measured.}
    \label{fig:scheme}
\end{figure}

Having analyzed the impact of edge-localized strain on the magnon band structure, we now investigate its implications for the magnon transport in finite nanoribbons under different strains and DMI scenarios. For this purpose, we consider a finite nanoribbon connected to two non-magnetic metallic leads as schematically illustrated in Fig. \ref{fig:scheme}. The injecting lead generates a spin accumulation that couples exclusively to the interfacial spins of the ZNR. This coupling is modeled as an effective magnetic field acting on the contact sites. Additionally, thermal fluctuations are included to account for the stochastic excitation of magnon modes. Magnons propagate along the ribbon toward the opposite side, where the spin current is detected at the ejecting lead once the system reaches a steady-state regime.

We start by describing the dynamics of the magnetic moments using the stochastic Landau-Lifshitz-Gilbert (LLG) \cite{landau1935theory,
gilbert2004phenomenological, Brown1963-nv} equations, associated with each localized spin state in our nanoribbon systems as $ \partial_t \bm{S}_i|_{\text{stoch}} = -\bm{S}_i\times \bm{h}^{\text{stoch}}_i$. The explicit expression becomes,
\begin{equation}
    \pd{\bm{S}_i}{t} = \mathbf{S}_i\times 
    \left[
        -\frac{\partial H}{\partial \mathbf{S}_i} 
        +\mathbf{h}^{\mathrm{stoch}}_i(t)
        -\frac{\alpha_i}{S}\,\pd{\S_i}{t}   
        +\frac{\alpha^{\mathrm{if}}_i}{S}\,\mathbf{S}_i\times\boldsymbol{\mu}_i
    \right] \,,
\end{equation}
where $S = |\S_i|$ and $H$ is the Heisenberg Hamiltonian from eq. \ref{eq:general_heis_ham11}. The total phenomenological Gilbert damping at site i is $\alpha_i = \alpha + \alpha^{\mathrm{if}}_i$,
where $\alpha$ denotes the site-independent damping and $\alpha^{\mathrm{if}}$ the interfacial  damping that is nonzero only in the contact regions between the nanoribbon and the leads \cite{Tserkovnyak2002-tq}. $\bm{\mu}_i$ describes the out-of-plane magnetic field that spins effectively feel in the nearby of the sites adjacent to the injecting lead. $\bm{h}^{\text{stoch}}_i(t)$ is the stochastic magnetic field that introduces thermal fluctuations by means of the dissipation-fluctuation theorem (FDT) and satisfies the conditions $\langle h^+_i(\omega) \rangle = 0$ and 
$\langle h^+_i(\omega) h^+_j(\omega') \rangle = 2\pi \delta_{ij}\delta(\omega-\omega') R_{ij}(\omega)$, where $h_i(\omega)$ is the Fourier transform of the circular component of the stochastic field  $h^+_i = h_i^x + ih_i^y$ and,
\begin{equation}\label{eq:Rw}
R_i(\omega) = 4(\omega - \mu_i)\frac{\alpha_i}{S}n_B(\omega-\mu_i) \,,
\end{equation}
encodes the thermal magnon population \cite{RuckriegelDuine2018-cx, Bender2017-ii, Brataas2015-yu}, where $n_B(\omega)$ is the Bose-Einstein distribution. The spin Hamiltonian $H$ is given by Eq. \ref{eq:lswt-ham}.

We look for the solutions that consist of low-energy excitations of the ground state configuration relying on LSWT, i.e., magnons. To this end, we assume small transverse (in-plane) components $S^x_i$ and $S^y_i$, corresponding to small deviations from the quantization axis. Under this assumption $S^z_i \approx S$ can be treated as constant, which allows to linearize the LLG equations. The resulting equation of motion for $S^+_i=S^x_i+iS^y_i$ is given by,

\begin{equation}\label{eq:linear-llg-eq}
\begin{split}
    i(1 + i\alpha_i) \pd{S^+_i}{t} &= 
    \bigl[S\sum_j (J_{ij} + \lambda_{ij}) + h^{\mathrm{ext}}\bigr] S^+_i \\
    &+\sum_j T_{ij}(\varepsilon)S_j^+
    + S h_i^+
    - i\alpha_i^{\text{if}}\mu_iS^+_i \,,
\end{split}
\end{equation}
where $T_{ij}$ is the strain-dependt matrix element defined in Eq. \ref{eq:lswt-ham}.

Transforming Eq. \ref{eq:linear-llg-eq} into frequency domain allows us to define the system's Green's function of the system, with the stochastic magnetic field acting as an external source, 
\begin{equation}\label{eq:Ginvij}
\begin{split}
G^{-1}_{ij}(\omega) =& \delta_{ij}\left[-(1+i\alpha_i)\omega +\sum_jS(J_{ij}+\lambda_{ij})+
i\alpha^{\text{if}}_iS\mu_i\right]  \\ & + T_{ij}\,,
\end{split}
\end{equation}
with $T_{ij}$ the matrix element of the LSWT Hamiltonian defined in Eq. \ref{eq:lswt-ham}. After normalizing the spin operator, the formal solution can be formulated in terms of this Green's function and the stochastic field $h^+_i(\omega)$ as the external source: 
\begin{equation}\label{eq:green-formal}
    \sum_j G_{ij}^{-1}(\omega)S^+_j(\omega) = h^+_i(\omega) \,.
\end{equation}

The spin current can be established by looking at the steady state regime of the system, which is governed by the expectation value of the equation of motion of the $S^z_i$ component. The steady state is achieved by the condition $\langle S^z_i \rangle = 0$. Substituting the formal solution Eq. \eqref{eq:green-formal} in  the equation of motion for the $S^z_i$ in Eq. \eqref{eq:linear-llg-eq}, we arrive at the final expression:
\begin{equation}\label{eq:llg-lswt}
\begin{split}
    &\langle \sum_j T_{ij}\text{Im}[S_i^-S^+_j]\rangle =\\
    &\quad-\langle {\alpha_i}\text{Im}[S_i^-\partial_tS_i^+]
    - \text{Im}[h_i^-S_i^+] - \alpha^{\text{if}}_i\mu_iS_i^+S_i^-\rangle\,.
\end{split}
\end{equation}
Here, the term on the left represents the spin current between nearest neighbors $i$ and $j$ mediated by the exchange interaction. On the right, we identify the spin current drain source due to the Gilbert damping term, alongside the injecting sources driven by the thermal heat bath, modeled by $h^+_i$, and the spin accumulation $\mu_i$ located at the contact regions between the ribbon and the leads. Based on Eq. \ref{eq:llg-lswt}, we define the aformentioned spin current terms acting on site $i$ as,
\begin{equation}
\begin{split}
   \sum_j\langle I_{i\rightarrow j} \rangle= \langle I^{\alpha}_i\rangle +
    \langle I^{\text{stoch}}_i\rangle + \langle I^{\mu}_i\rangle \,.
\end{split}
\end{equation}

Using the solution \ref{eq:green-formal} in the right side of  \eqref{eq:llg-lswt}, we find that the total injected spin current at the end of the nanoribbon as,
\begin{equation}\label{eq:Iw_final}
    I = \sum_{i\in \text{frontier}}\text{Im}\int\frac{d\omega}{2\pi}\, [STG(\omega)R(\omega)G^{\dagger}(\omega)]_{ii}\,,
\end{equation}
with $I$ the total integrated spin current at the ejection zone of the ZNR, $R(\omega)$ the encoding thermal fluctuations from the stochastic field \ref{eq:Rw}, $i$ the lattice positions where we measure the spin current, and $T$ the hopping matrix of the LSWT magnon model as in \ref{eq:lswt-ham}. $G(\omega)$ can be computed through the inversion of \eqref{eq:Ginvij}. The total ejected spin current can be separated by energy windows,hence, separating states in the magnon gap near the crossing point of the chiral bands from the bulk states located below that gap, we can study the behavior of topological edge states in spin transport and its response to the proposed strain.


\section{Effect of local edge strain in magnon dispersion}
\label{sec3}

\begin{figure*}[!t]
    \centering
    \includegraphics[width=\textwidth]{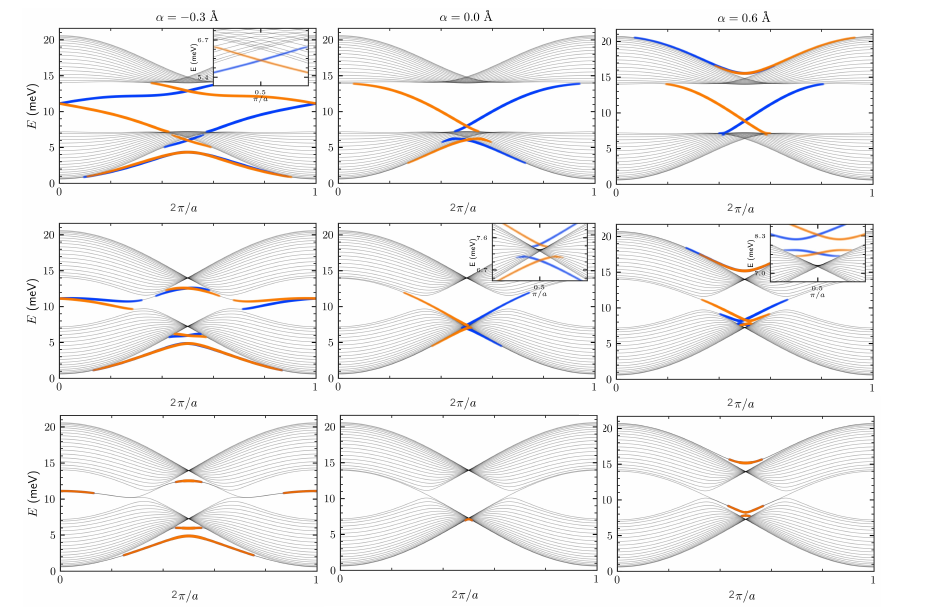}
    \caption{
    Magnon dispersion relations for $W=20$ ZN under varying strain parameter $\alpha$ ($\tau = a_0/4$) and $D=0\,\mev$ (bottom row), $D=0.1\,\mev$ (middle row) and $D=0.44\,\mev$ (bottom row), along k-path $k\in[0, \frac{\pi}{a}]$. The columns correspond to strain parameters $\alpha \in \{-0.3, 0, 0.6\}$\AA. Fixed parameters are $J = 2.2 \,\mev$, $\lambda = 0.1 \,\mev$, $h_{\text{ext}} = 0.1 \,\mev$ and $S = 3/2$. We have picked up the four outermost sites from both sides (orange: left, blue: right) of the ZNR and highlighted the resulting magnon state weights over the bands (fainted black solid lines). }
    \label{fig:bands_alpha_dmi}
\end{figure*}

 We investigate the strain-induced modifications of the magnetic exchange interactions by applying an edge-localized strain to the ZNR system. The ZNR is characterised by its width (W), defined as the number of zigzag chains across the transverse direction of the ribbon. The system is assumed to be periodic along the ribbon axis and finite along the transverse direction, resulting in a one-dimensional Brillouin zone for the longitudinal momentum (k). The band structure is obtained from the the LSWT Hamiltonian defined in eq. \ref{eq:Hij_momentum}. This strain profile is designed to tune the imbalance between edge and bulk coordination environments, thereby interpolating between the ideal LSWT honeycomb limit and the graphene-like edge-band regime. To this end, we employ the displacement field introduced in Eq.~\eqref{eq:interpolated-model}, with parameters fitted to the DFT results shown in Fig.~\ref{fig:xc_strain_dft}. 

Fig. \ref{fig:bands_alpha_dmi} shows the magnon dispersion relations resulting from the application of a non-uniform strain for increasing values of the DMI and a range of strain intensities $\alpha_{str}$. To distinguish between edge and bulk states, bands are highlighted in orange (blue) according to the magnon eigenstate weights projected onto the outermost A-sublattice (B-sublattice) and second-row B-sublattice (A-sublattice) sites of the left (right) edge. In particular, we use the eigenstate weights $|c_{n}(k)|^2 = \sum_{i\in \text{left (right)}} |M_{in}(k)|^2$. 

The central column corresponds to the case without strain, while the left and right columns show the dispersion relations for compressive and tensile strain respectively. Considering that second neighbor DMI values obtained from inelastic neutron scattering experiments of bulk CrI$_3$ are of the order of 0.3 meV\cite{Chen2018}, we have studied three different cases: trivial ZNR with $D = 0$ meV (bottom panel), a ZNR with $D = 0.1$ meV, a DMI smaller than the one predicted experimentally (middle panel), and a ZNR with $D = 0.44$ meV, a DMI slightly higher than the experimental one (top panel).

In the absence of strain ($\alpha = 0.0\,\text{\AA}$) and DMI ($D=0\,\mev$), there is a negligible presence of trivial edge states at $k \sim \pi$ (see small orange projection in the bottom central panel), which comes from the zigzag nature of the ribbon edge, very similar to the fermionic case\cite{Nakada1996, Wakabayashi_2010}. However, the imbalance in magnon on-site energies, which differ between edge and bulk sites due to the reduced coordination numbers at the edges, shift the energy of these states to lower values, around $\sim 7$ meV. In the case of a small DMI ($D = 0.1\,\mev$) a topological spin-wave gap between optical and acoustic magnon branches of $\Delta_\text{SW} = 1.85\,\mev$ appears. Here, the topological magnon gap refers to the value obtained by diagonalizing Eq. \ref{eq:Hij_momentum} for monolayer CrI$_3$.  The non-trivial edge states located at $E \sim 7$ meV for $D = 0$ meV, split into two copies with opposite chirality (see inset in the middle central panel). The lower energy chiral edge states come from the outer sites of the edge which present a lower coordination leading to a slightly smaller onsite energy. The chiral states with higher energy correspond to the inner edge sites which conserve the coordination number of the two-dimensional system. The top central panel shows the case with strong DMI ($D = 0.44\,\mev$), the magnon gap increases to $\Delta_\text{SW} = 6.68\,\mev$ and the non-trivial edge states with opposite chirality show a higher splitting. While the excitation energies of the lower pair lie at values similar to the lower bulk manifold, making difficult its characterization, the upper pair is completely isolated in the gap and shows a high linear dispersion.  

Applying compressive strain in the edge ($\alpha = -0.3\,\text{\AA}$ reduces the exchange interaction as shown in Fig. ~\ref{fig:xc_strain_dft}, lowering even more the on-site energy of the edge magnons. For $D = 0\,\mev$, this leads to the emergence of localized magnon edge states appearing at lower energies (see Fig. \ref{fig:bands_alpha_dmi}, bottom-left). These low energy states ($E < 7\,\mev$) are localized in the outer and inner sublattice edge sites respectively. In contrast to the $\epsilon = 0$ case, here the splitting corresponding to the pair of edge states is clearly visible due to the strain inducing a higher energy difference between the outer and inner sites. A quasi-flat band forms between the acoustic and optical magnon branches showing a strong weight in the inner edge sites close to the Brillouin zone boundaries ($k = 0$ and $k = 2\pi/a$). The higher energy band showing edge projection, shows a combination of inner and outer edge sites. For $D = 0.1\,\mev$ (middle right panel), the lower energy bands belonging to the outer edge sites do not show any chirality (orange and blue edge projections overlap). The other two bands become chiral with crossing points at $k = \pi/a$ and opposite chirality. Increasing the DMI ($D = 0.44\,\mev$), the magnon bands show a similar structure compared to the $D = 0.1$ meV case. However, the increase of the DMI induces a higher dispersion of the chiral edge states and increases the topological gap allowing the observation of the higher energy chiral edge bands (Fig. \ref{fig:bands_alpha_dmi}, top left panel).

In the tensile regime ($\alpha = 0.6\,\text{\AA}$), the exchange interaction is increased near the edge. However, since the tensile strain affect less the value of the Heisenberg exchange coupling between Cr atoms (see Fig. ~\ref{fig:xc_strain_dft}), it is expected that the magnon dispersion will be slightly modified with respect to the case without strain. For $D = 0\,\mev$ (Fig. \ref{fig:bands_alpha_dmi}, right bottom panel), the magnon dispersion shows three states around $k = \pi/a$ which are localized at the edges. In this case, the outermost sites contribute mainly to the central edge band, while the low-energy and high energy bands have contributions from the inner and bulk sites. For $D = 0.1\,\mev$, the first two edge bands become chiral, while the third one at high energy shows no chirality. The combination of tensile strain and higher DMI (Fig. \ref{fig:bands_alpha_dmi}, top right panel), gives rise to a pair of chiral localized edge states completely isolated inside the gap coming from the outer edge sites. The higher energy edge state becomes chiral only at energies higher than $17\,\text{meV}$ (for lower energies left- and right-moving magnons are localized in both edges). The low energy chiral edge state, visible for $D = 0.1\,\mev$, hybridizes completely with the bulk states and disappear.

\begin{figure}[!t]
    \centering
    \includegraphics[width=\linewidth]{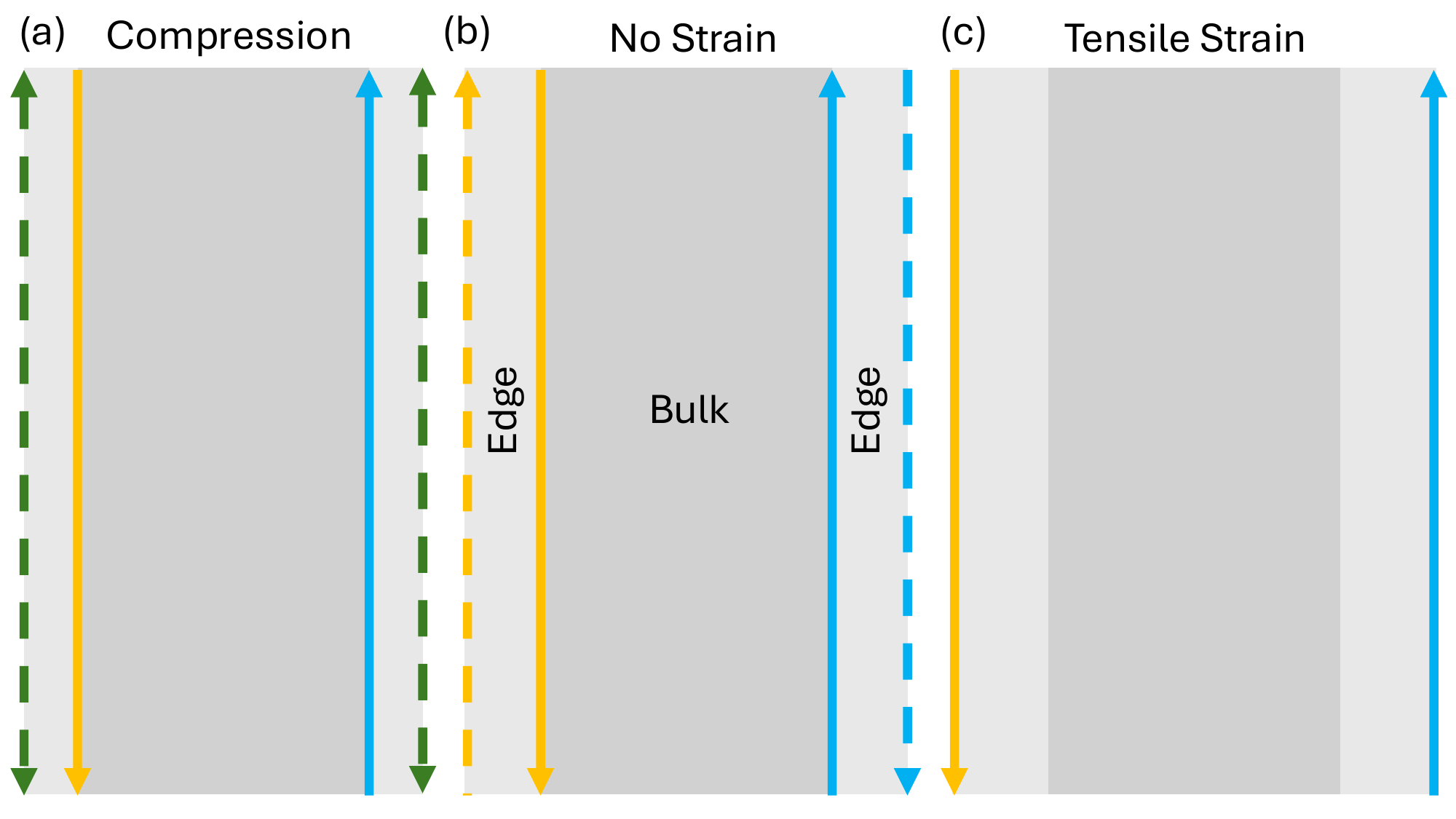}
    \caption{Schematics of the magnon edge state formation for  strain and unstrained CrI$_3$ nanoribbons. The light gray regions represents the edge, containing inner (close to the bulk) and outer sites. The dashed arrows correspond to the lower energy magnon edge bands, while the normal arrows correspond to the second magnon edge state. The color of the arrows are the same that the ones used to plot the band dispersion. We have used green to denote a non-chiral edge magnon.}
    \label{fig:edgestates}
\end{figure}

\section{Magnon-mediated spin transport in CrI$_3$ nanoribbons}
\label{sec4}

After analyzing the impact of the local strain on the magnon band structure, we focus now on the study of its impact on the magnon transport properties by considering the device depicted in Fig. \ref{fig:scheme} for a nanoribbon with fixed width $W=8$. We investigate the decay of spin currents by varying the ribbon length $l$. In all simulations, Anderson disorder is modeled by incorporating randomly distributed large on-site energy perturbations with a concentration of $w_{\text{deff}}=15\%$ relative to the total number of lattice sites. The spin currents are computed using the relaxation LLG-LSWT formalism (Eq.~\eqref{eq:Iw_final}), where the Green's functions are constructed from the LSWT Hamiltonian incorporating the strain model defined in Eq.~\eqref{eq:interpolated-model}. We focus specifically on the strong DMI regime ($D=0.44\,\text{meV}$), as our previous band structure analysis (Fig. \ref{fig:bands_alpha_dmi}) demonstrated that this regime effectively isolates the topological edge states within the band gap and show higher band dispersion.

\begin{figure}[t]
    \centering
    \includegraphics[width=1\linewidth]{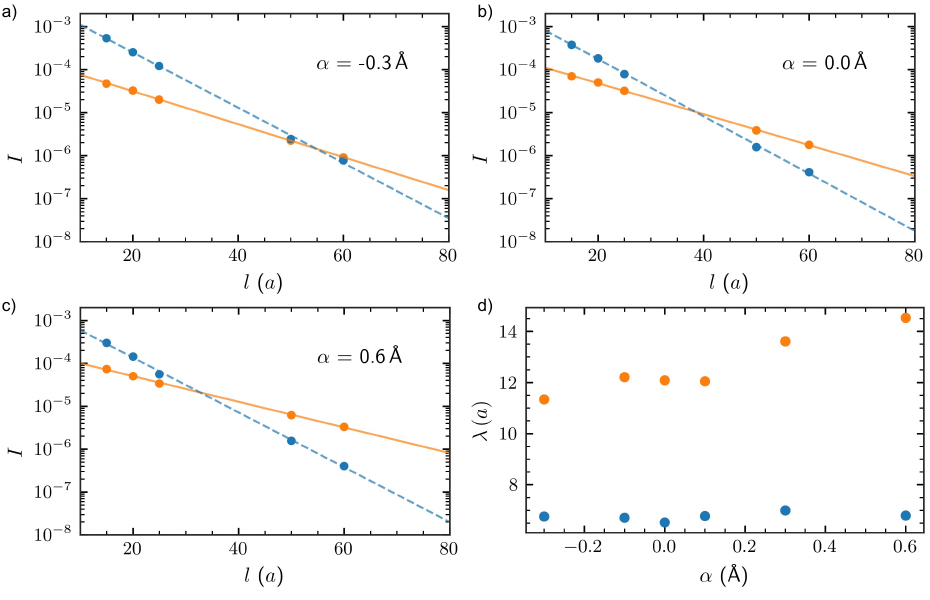}
    \caption{Spin currents ejected as a function of the ribbon length $l$ and strain amplitude $\alpha$. The simulation parameters are fixed at $J=2.2\,\text{meV}$, $J_{\text{ani}}=0.1\,\text{meV}$, $D=0.44\,\mev$, $h_{\text{ext}}=0.1J$, $T=0.8J$, with a defect concentration of $15$\%. (a)-(c) Spin currents for compressive ($\alpha=-0.3$\,\AA), zero, and tensile ($\alpha=0.6$\,\AA) strain. Orange (blue) markers represent the edge (bulk) spin current contributions. Dashed lines indicate the exponential fits $I(l) \propto e^{-l/\lambda}$. (d) Characteristic decay length $\lambda$ as a function of the strain parameter $\alpha$, extracted from the fits.}
    \label{fig:spincurr1}
\end{figure}

Figure \ref{fig:spincurr1}(a-c) presents the magnitude of the calculated edge (orange) and bulk (blue) spin currents $I$ as a function of length for three representative strain amplitudes. As expected for a disordered system, we observe a clear exponential suppression of the spin current, following $I(l) \propto e^{-l/\lambda}$. Notably, the edge contribution consistently dominates the transport, exhibiting a slower decay than the bulk, which confirms the topological protection of the edge modes even in the presence of significant disorder.

A comparison of the panels shows that the decay length is highly sensitive to the applied strain. This dependence is quantified in Fig.~\ref{fig:spincurr1}(d), where the characteristic decay length $\lambda$ is plotted as a function of the strain amplitude $\alpha_{\text{str}}$. The data exhibit a monotonic increase of $\lambda$ with $\alpha_{\text{str}}$: under compressive strain ($\alpha<0$) spin transport is strongly suppressed, producing short edge decay lengths ($\lambda_{\mathrm{edge}}\sim 11a$), whereas tensile strain ($\alpha>0$) extends the transport range, raising $\lambda_{\mathrm{edge}}$ to $\sim 14a$ at $\alpha=0.6$\,\AA. As for the bulk currents, strain does not have a great impact on its decay lengths $\lambda_{\mathrm{bulk}} \sim 7.5a$.
To explain this behavior, we need to compare with the magnon band dispersions obtained in the previous section. 
It is important to note that low-energy magnon states contribute significantly more to the overall spin transport than high-energy ones. This occurs because the thermal occupation of high-energy modes is exponentially suppressed at lower temperatures, as dictated by the Bose-Einstein  distribution (see Supplemental Material \cite{supplementary} for further details about the thermal modulation of occupations.)
For this reason we focus on bands with $E < 10\,\text{meV}$. We see a clear trend, namely, the decay length of the spin currents mediated by topological magnons increases as the crossing point (or critical point) of the topologically-protected magnon edge bands shift towards the gap. This clearly pinpoints the importance of edge strain to decouple edge transport and bulk transport in order to maximize spin currents in 2D ferromagnetic insulators. 

\section{Conclusions}
\label{sec5}

In this work, we have theoretically investigated the manipulation of magnon transport in monolayer CrI$_3$ zigzag nanoribbons through the application of edge-localized strain. By combining first-principles DFT calculations with Linear Spin Wave Theory (LSWT) and Green's function transport simulations, we have established a direct link between mechanical deformation, magnetic exchange modulation, and the robustness of spin currents. We have demonstrated that mechanical deformation can control the localization of the topological magnon states at the edges, thereby providing a new platform for manipulating localization in specific sublattices and shifting these states to lower or higher energies.

In the trivial DMI regime ($D=0$), we showed that compressive strain restores localized edge states that are otherwise absent due to the edge-bulk energy mismatch. Conversely, in the topological regime (finite DMI), strain acts as a powerful control knob for the chiral edge modes. We found that while compressive strain tends to hybridize the topological states with the bulk manifold, tensile strain lifts the chiral edge states allowing to isolate them within the gap, allowing for a potential characterization.

Finally, our transport calculations in disordered ribbons provided quantitative evidence of these topological protection mechanisms. We observed that the characteristic decay length of the spin current ($\lambda$) is strongly modulated by the applied strain field. Specifically, tensile strain significantly enhances the transport range ($\lambda \approx 14a$) by protecting the chiral edge channels from scattering into the bulk. In contrast, compressive strain reduces the decay length ($\lambda \approx 11a$) by promoting edge-bulk hybridization.

These findings highlight the potential of magnon straintronics as a viable route for engineering robust, dissipationless spin transport channels in two-dimensional magnetic materials. The ability to mechanically tune the topological properties of magnons without altering the chemical composition paves the way for developing flexible and energy-efficient spintronic devices.

\bibliography{biblio.bib}

\appendix

\end{document}